\documentclass[pre,amsmath,showpacs]{revtex4}
\usepackage{graphicx}
\usepackage{color}
\usepackage{bm}
\usepackage{epsfig}
\usepackage{latexsym}
\usepackage{float}
\usepackage[utf8]{inputenc}
\graphicspath{ {figure/} }

\begin{document}

\newcommand{\be}{\begin{equation}}
\newcommand{\ee}{\end{equation}}
\newcommand{\bea}{\begin{eqnarray}}
\newcommand{\eea}{\end{eqnarray}}
\newcommand{\nn}{\nonumber}


\title{Actin filaments growing against a barrier with fluctuating shape}
\author{Raj Kumar Sadhu and Sakuntala Chatterjee}
\affiliation{Department of Theoretical Sciences, S. N.
Bose National Centre for Basic Sciences, Block  JD, Sector  III, Salt Lake,
Kolkata  700106, India. }


\begin{abstract}
We study force generation by a set of parallel actin filaments growing against a non-rigid obstacle, in presence of an external load. The filaments polymerize by either moving the whole obstacle, with a large energy cost, or by causing local distortion in its shape which costs much less energy. The non-rigid obstacle also has local thermal fluctuations due to which its shape can change with time and we describe this using fluctuations in the height profile of a one dimensional interface with Kardar-Parisi-Zhang dynamics. We find the shape fluctuations of the barrier strongly affects the force generation mechanism. The qualitative nature of the force-velocity curve is crucially determined by the relative time-scale of filament and barrier dynamics.  The height profile of the barrier also shows interesting variation with the external load. Our analytical calculations within mean-field theory show reasonable agreement with our simulation results.        
\end{abstract}
\maketitle

\section{Introduction}

Cell motility plays an important role in a wide variety of biological processes like morphogenesis, wound healing or tumor invasion \cite{review1,review2,review3, review4}. Actins and microtubules are cytoskeletal proteins whose polymerization and depolymerization can generate significant forces, without any assistance of molecular motors, and propel the cell forward. In presence of a biological barrier, these filaments elongate and generate a pushing force against the barrier and in many in vitro studies this force has been measured explicitly by applying an external load on the barrier in the opposite direction. With increasing load, the velocity of the barrier decreases and the functional nature of dependence of velocity on the applied force is an important characteristic of the force generation mechanism. The maximum polymerization force generated by the filaments is known as `stall force' and is measured as the minimum load required in order to stall the barrier motion completely. There has been a surge of experimental as well as theoretical research activities to determine the stall force and the force-velocity characteristic of the cytoskeletal filaments in the last few years.

Interestingly, the qualitative nature of the force-velocity curve was found to
depend on the details of the experimental set-up. A convex force-velocity 
characteristic was reported for actin quoted polystyrene beads \cite{marcy2004} 
and magnetic colloidal particles pushed by unbranched parallel actin filaments 
\cite{baudry2011,baudry2014}. On the other hand, a concave force-velocity curve 
was obtained for branched actin network \cite{theriot2005}, where
velocity remains almost constant for small load and drops rapidly at large load. 
An even more complex force-velocity relationship was measured for lamellipodial 
protrusion in a keratocyte, where velocity showed rapid decay for very small 
load, followed by a plateau at moderate load and another rapid decay close to 
stalling \cite{mogilner2006,zimm}. Although multiple filaments are expected to 
generate larger force than single filament \cite{theriot2003,marcy2004,mogilner2006}, in  
\cite{theriot2007} the stall force of  approximately eight actin filaments was 
measured and found to be in the piconewton range, close to a single filament 
stall force \cite{pollard2004}, indicating absence of co-operation among the 
filaments.

To investigate the force-velocity relationship theoretically, several different
models have been proposed. Force generation by a single actin filament growing
against a barrier has been explained using a simple Brownian ratchet mechanism
where thermal fluctuations of the barrier creates a gap between the barrier and
the filament tip, making it possible for the filament to grow by adding one
monomer in the gap \cite{peskin1993}. This mechanism predicts a convex
force-velocity curve. This simple model has been subsequently generalized where 
details of interaction between the monomers and the barrier has been considered 
\cite{carlsson2000} and flexibility of the filament has been included
\cite{jphys2006}. In all these cases existence of a convex force-velocity
relationship has been verified. However, when the  Brownian ratchet mechanism
was extended for multiple filaments, the nature of the force-velocity curve was
found to crucially depend on how the details of the interaction and load-sharing
among the filaments were modeled \cite{schaus,krawczyk2011,kirone,ddas2014}. 
Certain models even showed a crossover from convex to concave force-velocity 
curve, as some model parameters are varied 
\cite{mogilner2012,hansda2014,carlsson2014}.

Inside a cell, actin filaments grow against the plasma membrane which is not a 
rigid object but elastically deformable \cite{nirgov}. 
Even in vitro, when the filaments 
push against an obstacle as they polymerize, the obstacle may in general have 
local shape deformations. 
In \cite{atilgan} a flexible plasma membrane was explicitly modeled
and it was shown that thermal fluctuation of this flexible obstacle
substantially enhances the growth velocity of a filopodial protrusion. It was
argued that in the case of a flexible membrane, a filament only has to overcome
the local bending energy in order to polymerize (whereas for a rigid obstacle
the full load must be overcome) and this gives rise to a larger
velocity for a given load. Effect of a flexible plasma membrane on actin network
growth was experimentally demonstrated in \cite{liu} when reconstituted actin
networks in vitro were assembled onto synthetic lipid bilayers and it was found
that the membrane elasticity causes formation of bundled filament protrusion 
from branched filament networks.

Motivated by this, we carry out a study to probe the detailed quantitative
aspects of interaction between a set of growing filaments and an obstacle whose
position as well as shape can fluctuate with time. To keep our description
simple, we model the obstacle by a one dimensional non-rigid object whose local 
thermal fluctuations can alter its shape and using a lattice gas model,
we describe it by a Kardar-Parisi-Zhang (KPZ) interface \cite{kpz}. In presence 
of an external load, the obstacle tends to move in the direction opposite 
to that of polymerization. In order to polymerize, the filaments must push against 
the barrier, either causing a local change in its profile 
(which requires less energy) or causing a global movement of the whole barrier
 (which involves a large energy cost). We are interested to find out how presence 
of the fluctuating  barrier affects the dynamics of the actin filaments, and how 
the presence of the filaments affects the shape of the barrier.

Our numerical simulations and analytical calculations show that there is a rich 
interplay between the polymerization dynamics of the filaments and the shape
fluctuations of the barrier. For small and intermediate values of the external 
force, the barrier motion is governed by its global movement, and for large
force, the local fluctuations become important. These local movements cost
less energy and can continue even when the force is significantly large. As a 
result, the stall 
force in our system is much higher than that for a rigid barrier \cite{kirone}. 
Moreover, these local movements may be caused by filament polymerization or by
independent thermal fluctuations of the barrier and hence the stall force may
also depend on the properties of the barrier. Indeed for a single
filament, the stall force is found to increase with the size of the barrier. 
For $N$ filaments stall force is independent of the barrier size and 
scales linearly with $N$. The barrier shape is also affected by the growing 
filaments and the scaling behavior of its height profile shows continuous
variation as a function of the external load.

There are two time-scales in our system, one associated with the (de)polymerization 
of the filaments and the other with the thermal fluctuations of the barrier. 
Our results show that the choice of these time-scales may crucially determine the 
nature of the force-velocity curve. This is because the local movements of the
barrier make increasingly important contribution to its velocity as the thermal 
fluctuations become faster. Even for small or intermediate load, therefore,
the barrier
velocity is not governed by its global movement alone and this changes the
qualitative nature of dependence of velocity on load.
The stall force is also found to decrease for faster barrier dynamics.

This paper is organized as follows. In section \ref{sec:model}, we describe our 
model. Our results for the single filament and multiple filaments are presented 
in sections \ref{sec:single} and \ref{sec:multi}, respectively, and conclusions 
are in section \ref{sec:sum}. 

\section{Description of the model}
\label{sec:model}

\begin{figure}
  \includegraphics[scale=0.8]{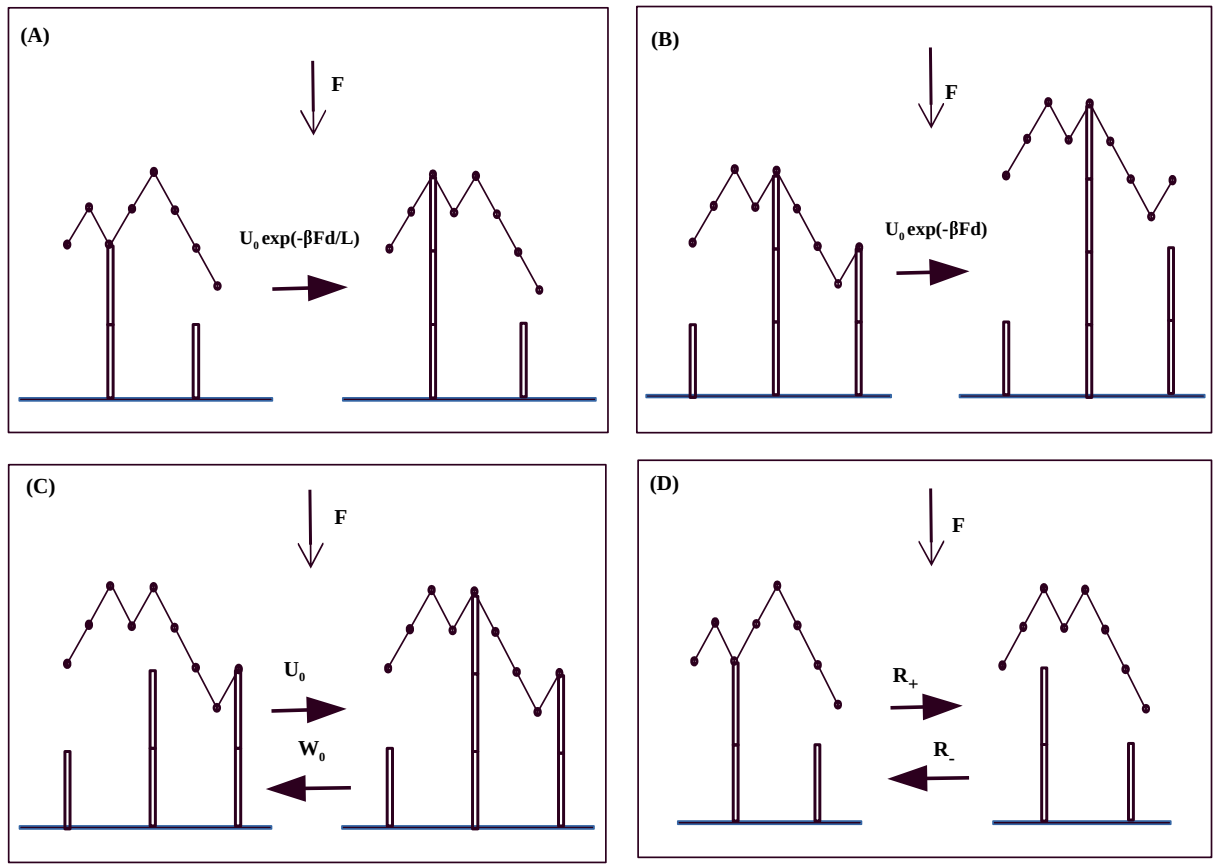} 
\caption{Schematic representation of our model. (A): Polymerization of 
a bound filament by causing a local change in barrier height with rate 
$U_0e^{-\frac{\beta Fd}{L}}$. (B): A bound filament polymerizes by 
causing global movement of the whole barrier with rate $U_0e^{-\beta Fd}$. 
(C): A free filament polymerizes and depolymerizes with rates $U_0$ and $W_0$,
respectively. Since these processes do not involve any barrier motion, these
rates are independent of $F$. (D): Thermal fluctuation of the barrier: 
a local valley can flip to a hill with rate $R_+$ and the reverse process occurs
with rate $R_-$. We use local detailed balance, $R_+/R_- =\exp(-\beta F d /L)$,
except at the binding sites, where hill to valley transition may be blocked due
to the presence of a filament.}
\label{model}
\end{figure}

Our model consists of $N$ parallel filaments growing against a barrier with a
fluctuating height profile (see Fig. \ref{model}). We model the filaments as 
rigid polymers, made of rod-like monomers of length $d$, such that a 
(de)polymerization event (decreases) increases the length of the filament by an 
amount $d$. The barrier is modeled as a one dimensional surface.  
In our lattice model, the discrete surface elements are represented as lattice 
bonds of length ${\lambda}$, which can have two possible orientations, 
$\pm \pi /4$. We 
denote these two cases by symbols $/$ and $\backslash$ and call them upslope and
downslope bonds, respectively. Height at any particular lattice site $i$ is
defined as $h_i = \delta/2 \sum_{j=1}^{i-1} \tan \theta_j$, where $\theta_j$ is
the orientation of the $j$-th bond and $\delta=\sqrt{2}{\lambda}$. The total 
number of such bonds is $L$. 
One $/$ followed by a $\backslash$ forms a local hill and in
the reverse order $ \backslash /$ they form a local valley. The local height of 
the surface fluctuates due to transition between these hills and valleys.
When a local hill (valley) at a given site flips to a valley (hill), the height 
of that particular site decreases (increases) by an amount $\delta$. We assume
$\delta$ is equal to the monomer length $d$. As explained below, this assumption
means that height fluctuation of the surface creates a gap which is just enough
for insertion of a monomer. Towards the end of the paper, we briefly discuss the
case of $\delta \neq d$. 

A filament whose tip is in contact with the barrier, is called a bound filament 
and in the absence of any such contact, it is called a free filament. The surface
site where a bound filament can form a contact, is called a binding site. 
When a bound filament polymerizes, it creates space for insertion of another monomer
by pushing the barrier up and in this process performs work against the external
load (which tends to push the barrier down). When the bound filament pushes against a 
local valley, that valley flips to a hill and the height of the binding site increases by an amount $d$ (Fig. \ref{model}A). However, polymerization of a bound filament, which is not in contact with a local valley, requires a global movement of the whole barrier, as 
shown in Fig. \ref{model}B, when height of all the $L$ sites are increased by an 
amount $d$. Assuming $F/L$ is the load per site, the energy cost for the first 
process is just $Fd/L$, and for the second process it is $Fd$. Following the 
rule of local detailed balance, we assign rates $U_0 exp(-\beta F d/L)$ and 
$U_0 exp(-\beta F d)$ to these two types of polymerization processes, respectively. 
Here, $\beta$ is the inverse temperature and $U_0$ is the free filament polymerization 
rate that does not involve any barrier movement and hence is independent of $F$. 
We also assume the depolymerization rate is same for both free and bound filaments 
and is denoted as $W_0$. When a bound filament depolymerizes, it loses contact 
with the barrier and becomes a free filament. In certain configurations, when 
there is only one bound filament, its depolymerization results in an unsupported 
barrier.

Apart from being pushed by the filaments, the barrier can also show thermal
fluctuations, when local hills can flip to valleys and vice versa. However, due
to presence of the filaments, these transitions can sometimes get blocked. For
example, if a bound filament is in contact with a hill, then that particular
hill cannot flip to a valley, until the filament depolymerizes and a gap is
created for a local downward movement of the barrier. When both forward and
reverse transitions are allowed, their rates rates satisfy local detailed balance
$\dfrac{R_{+}}{R_{-}}=e^{-\beta Fd/L}$, where $R_+$ is the rate at
which local surface height can increase (i.e. a valley flips to a hill) and
$R_-$ be the reverse transition rate. Note that in the absence of any external load
$F$, the transition between hills and valleys become symmetric at all sites
other than the binding sites and the surface has a local Edwards-Wilkinson  
dynamics \cite{ew}. For non-zero $F$, hill to
valley transitions are generally favored (except, possibly, at the binding
site) and the barrier behaves like a KPZ surface with a downward bias.

We assume periodic boundary condition for the surface and an equal number of
upslope and downslope bonds, i.e. no overall tilt. In one Monte Carlo step, we
attempt to perform $N$ filament updates (polymerization or depolymerization) and 
$S$ independent (unaided by the filaments) surface updates. By changing the value of
$S$ we can tune the relative time-scale between filament dynamics and barrier
dynamics. For smaller (larger) $S$ value, the barrier dynamics is slower (faster)
than the filament dynamics. A relative time-scale between the surface and filament
dynamics can also be introduced by rescaling $R_+$ and $R_-$, but we have used
$R_-=U_0$ and $R_+=U_0 e^{-\beta Fd/L}$ throughout and controlled the relative
time-scale by $S$ instead. We start with an initial configuration where all $N$
filaments have unit length, containing one monomer each and  the upslope and
downslope bonds are placed alternatingly (a flat surface). We let the system 
evolve for a long time, according to above dynamical rules. All our measurements
are performed in the steady state.

\section{Results for single filament}
\label{sec:single}

\begin{figure}
\includegraphics[scale=0.7]{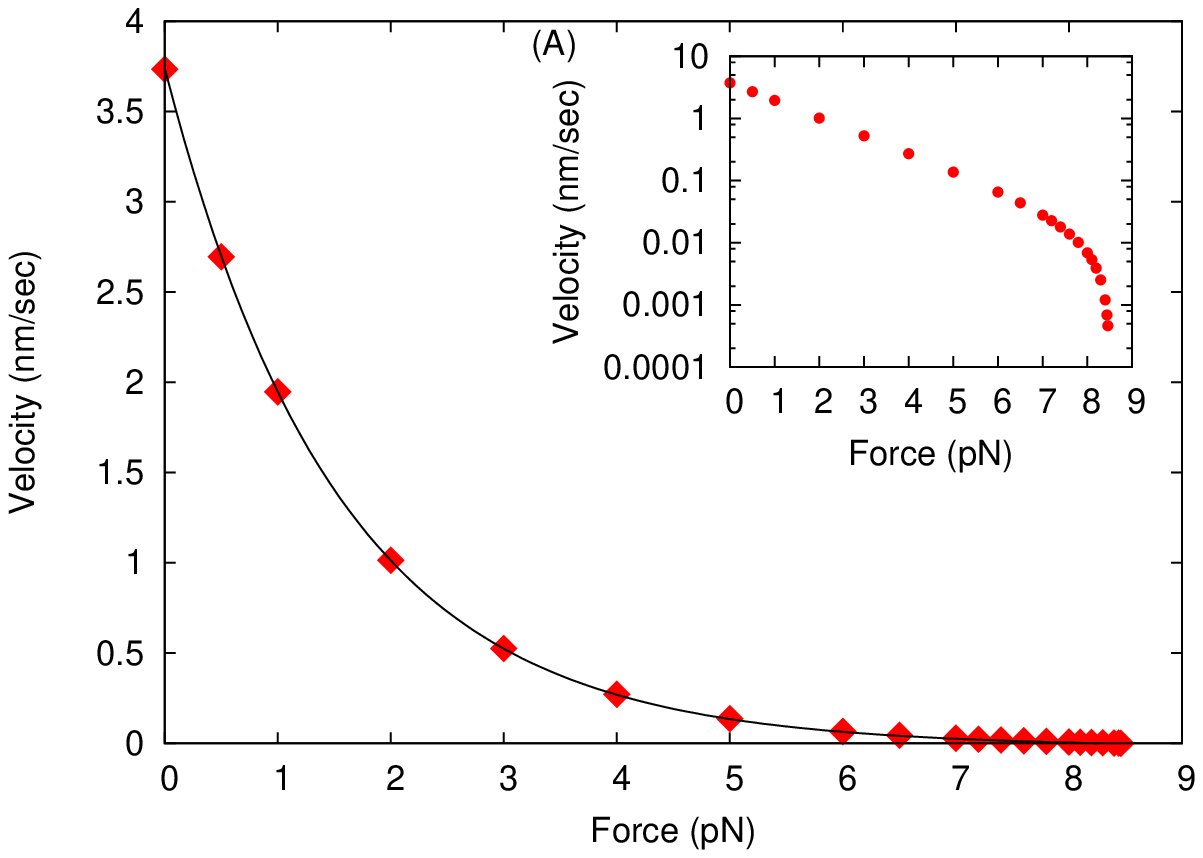} 
\includegraphics[scale=0.7]{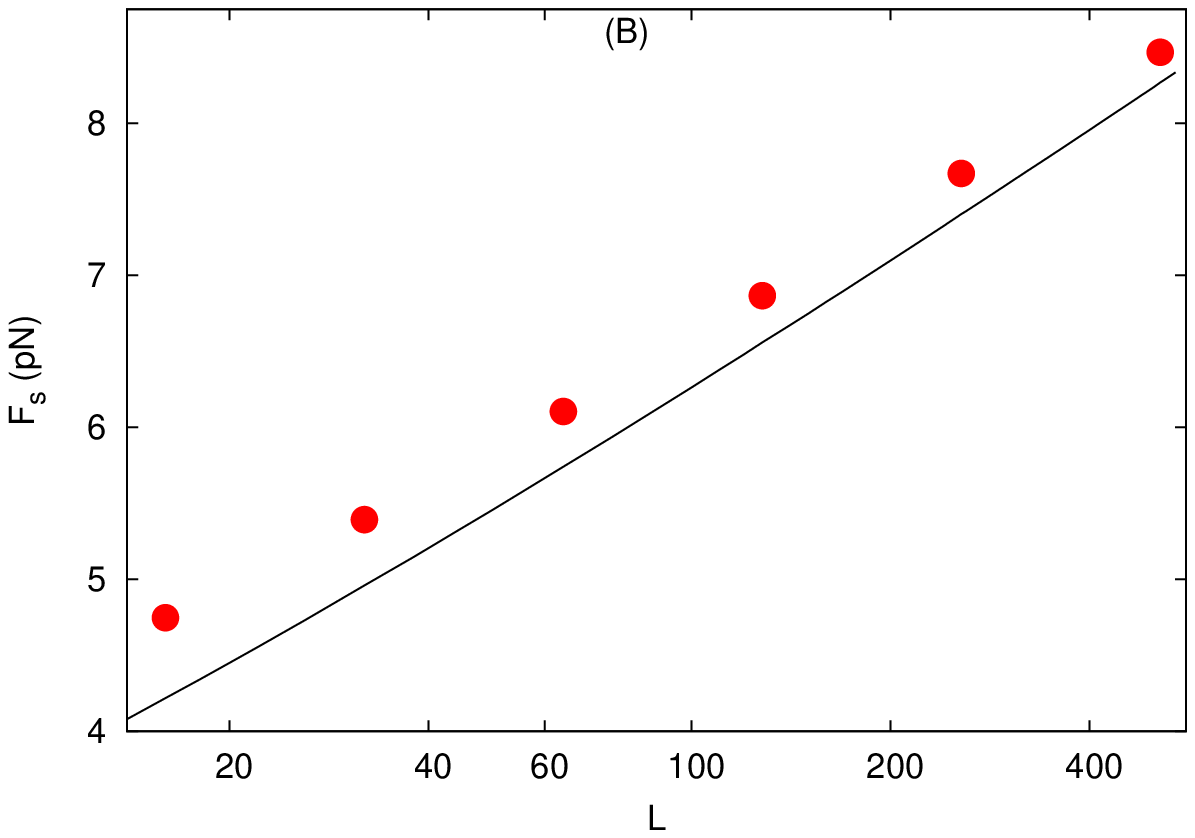}
\caption{Force-velocity characteristic and stall force for a single filament. 
(A): Force-velocity curve has a convex
shape. Inset shows exponential decay of the barrier velocity for small and
intermediate $F$, when the global motion of the barrier dominates. Close to
stalling the local fluctuations become important. We have used $L=512$ here.
 (B): Stall force increases with the barrier size $L$. In both the panels, we have 
used $S/L=1$. The 
free filament depolymerization rate $W_0=1.4 ~s^{-1}$ \cite{pollard,review1} and the
polymerization rate $U_0$ is proportional to the free monomer concentration 
with a proportionality constant $k_0=11.6 ~\mu m^{-1}s^{-1}$ \cite{pollard,review1}.  
We have used a monomer concentration $C = 0.24 ~\mu m$, which gives 
$U_0=2.784 s^{-1}$. The monomer size is $d=2.7 ~nm$ \cite{review1,hansda2014}. 
At room temperature the parameter $\beta d=0.65 ~pN^{-1}$. Discrete points 
show simulation data and continuous lines show analytical results.}
\label{F_vs_V}
\end{figure}

For a single filament, we first present the results for $S=L$ and later we
consider the effect of variation of $S$. We define the velocity $V$ of the barrier
as the rate of change of the average height of the surface after the system has
reached steady state. We present the force-velocity curve in Fig. \ref{F_vs_V}A.
This curve has a convex shape where velocity decays rapidly for small force, and 
for large force it decays slowly. In fact for small and intermediate values of
force, the velocity falls off exponentially (Fig. \ref{F_vs_V}A inset) and close
to stalling it shows deviation from the exponential form. We explain below
that the exponential dependence originates from the global movement of the
barrier (as shown in Fig. \ref{model}A) which dominates $V$ for small and
moderate $F$ range. In Fig. \ref{F_vs_V}B we show the variation of stall force 
$F_s$ with the barrier size $L$. Stall force increases with $L$, although 
logarithmically slowly. 
Note that the stall force is often interpreted as the maximum polymerization force
generated by the filament and therefore it is somewhat surprising that it
depends on the size of the barrier. We show below that in our system the local fluctuations of the barrier, which depend on $L$,
make substantial contribution towards its net
velocity and this becomes particularly significant in the stalling regime. 

In our system there are two possible barrier movements: global and local. In a
global movement, a bound filament polymerizes by pushing the whole barrier up,
such that the average height changes by an amount $d$. The rate at which this
process happens is $U_0 \exp( -\beta F d)$. Let this process contribute a velocity 
$V_1$ to the barrier in the steady state, which can be written as  
\be
V_1 = p_0 d U_0 exp(-\beta F d).
\ee
Here, $p_0$ is the probability that the filament is in
contact with the barrier. Note that here
we have ignored the possibility that the bound filament is pushing against a
valley (in that case no global movement takes place, only a local flip is
sufficient for polymerization). In fact we have verified in our simulation (data
presented in Fig. \ref{N_1_density_profile}B ) that the probability of finding  
a valley at the binding site is indeed small.

To write $V_1$ as a function of $F$ we still need to calculate $p_0$. Define
$p_i$ as the probability that the distance between the filament tip and the
binding site is $i$. Clearly, $i=0$ corresponds to
 the contact probability. It is easy to 
see that for $i>0$, the probability $p_i$ satisfies master equation for a biased
random walker:  
\be
\dfrac{dp_i}{dt}=W_0p_{i-1}+U_0p_{i+1}-(W_0+U_0)p_i
\ee
and for $i=0$ one has 
\be
\dfrac{dp_0}{dt}=U_0p_{1}-W_0p_0.
\ee
 Here, we have 
ignored any change in $p_i$ due to height fluctuations at the binding site. For 
fast barrier dynamics, when height fluctuations increase, this assumption 
breaks down. In the steady state, these
equations yield a recursion relation $p_i=\Big(\frac{W_0}{U_0}\Big)^i p_0$  
for positive $i$. This recursion relation, along with the normalization condition
$\sum_i p_i =1$ yields the expression $p_0=(1-W_0/U_0)$, which is independent of
$F$. So the final expression for $V_1$ becomes 
\be
V_1 = d (U_0 - W_0) \exp( -\beta F d).   
\ee 
To calculate the velocity due to local height fluctuations of the barrier, we
consider a local valley (hill) flipping to a hill (valley) which increases
(decreases) the average height by an amount $d/L$. As discussed in section 2, 
the transition rates at the binding site is different from the rest 
of the system, since a hill to valley transition may be blocked, if a filament 
is in contact. Then the barrier velocity due to local height fluctuations can 
be written as
\be
V_2= \frac{d U_0}{L} \left [ \left ( (1+p_0) p_v(0) + \sum_{i=1}^{L-1} p_v(i)
\right ) e^{-\beta F d/L} - (1-p_0) p_h(0) - \sum_{i=1}^{L-1} p_h(i) \right ]
\ee
where $p_v(i)$ and $p_h(i)$ denote the probabilities to find a valley
and a hill, respectively at a distance $i$ from the binding site. In the above
equation, the first term on the right-hand-side represent the situation
where a valley at the binding site flips to a hill, due to thermal fluctuations
or due to being pushed by the filament. The second term present flipping of a
valley to a hill at all the other sites. The third term describe the case when
there is a hill at the binding site which can flip to a valley when no filament
is in contact. The fourth term describe flipping of a hill to a valley in rest
of the system. The probabilities $p_v(i)$ and $p_h(i)$ 
can be calculated within a mean field approximation by considering
a KPZ surface with the binding site acting as a `defect site' (see Appendix \ref{app:toy} for details), where the transition rates are different from the
rest of the system. Our calculations show that $p_v(i)$ and $p_h(i)$ have a
rather weak dependence on $F$ and their difference $[p_v(i)-p_h(i)]$ is independent
of $i$ and scales as $1/L$. For large $L$, the total velocity of the barrier 
$V=V_1+V_2$ can be written as 
\be
V(F) = d(U_0-W_0)~e^{-\beta F d} + \frac{d U_0}{L} \left [ p_v(0) (1+p_0) -(1-p_0)
p_h(0)+  \sum_{i=1}^{L-1} \{ p_v(i)(1-\frac{\beta d F}{L})-p_h(i) \}  \right ]
\label{eq:vf}
\ee
where we have retained terms upto order $1/L$ and ignored higher order terms.  
In Fig. \ref{F_vs_V}A we compare our calculation with simulation results and
obtain reasonably good agreement. For small $F$, the first term in Eq. \ref{eq:vf} dominates 
the velocity and as $F$ increases, local fluctuations become more important. The last term in 
Eq. \ref{eq:vf}, within the braces, which represents the velocity due to hill-valley 
fluctuations at all sites, except the binding site, is the most dominant term in the local
movement. In the stalling region, the positive contribution from the global movement and the negative contribution from the local fluctuations cancel each other, where the first and last terms of Eq. \ref{eq:vf} determine the major balance. The stall force $F_s$ can be obtained by 
graphically solving the above transcendental equation after putting its left hand side zero. 
This gives stall force as a function of $L$ and we compare this variation with
simulation results in Fig. \ref{F_vs_V}B. We find good agreement for large $L$
but as expected, for small $L$ there are deviations. Note that the stall force 
in our system is substantially higher than that for a rigid barrier \cite{kirone}. Since the local movements cost much less energy, they can continue even when the load is high.


\subsection{Effect of faster and slower barrier dynamics}

\begin{figure}
\includegraphics[scale=0.7] {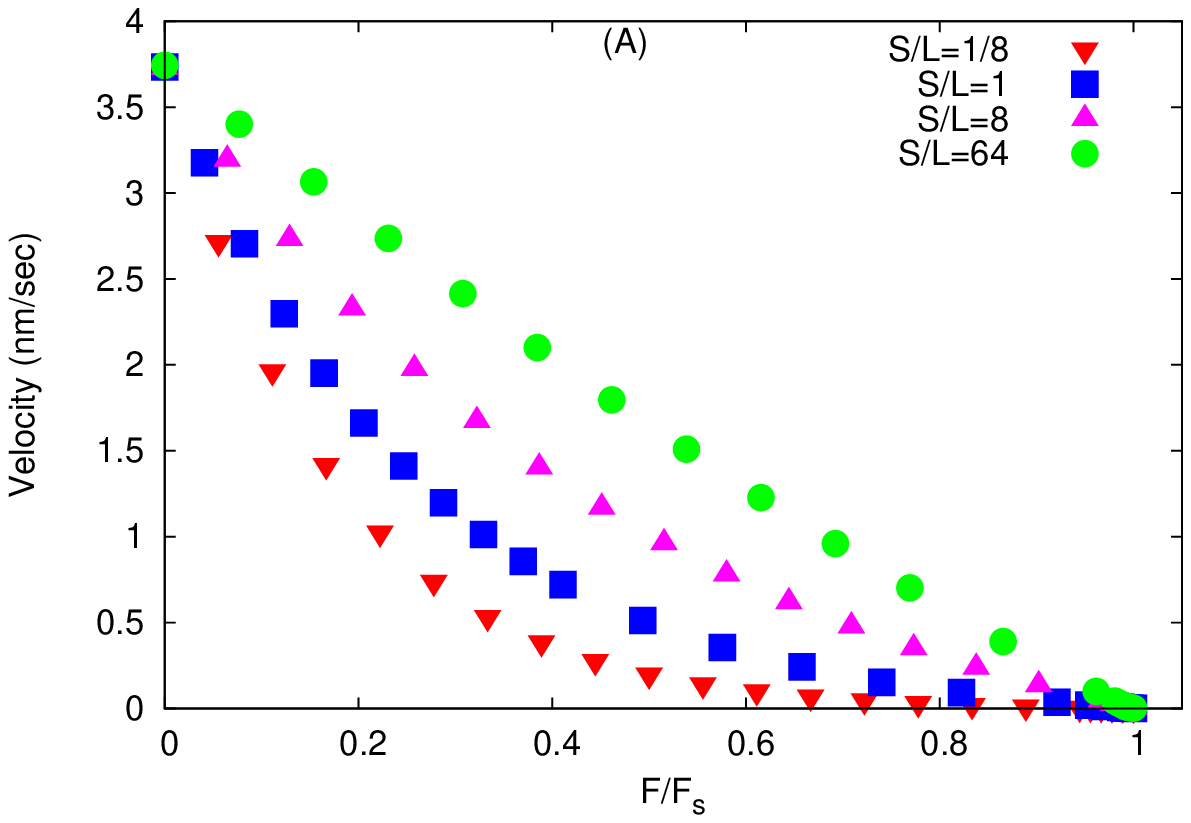}
\includegraphics[scale=0.7] {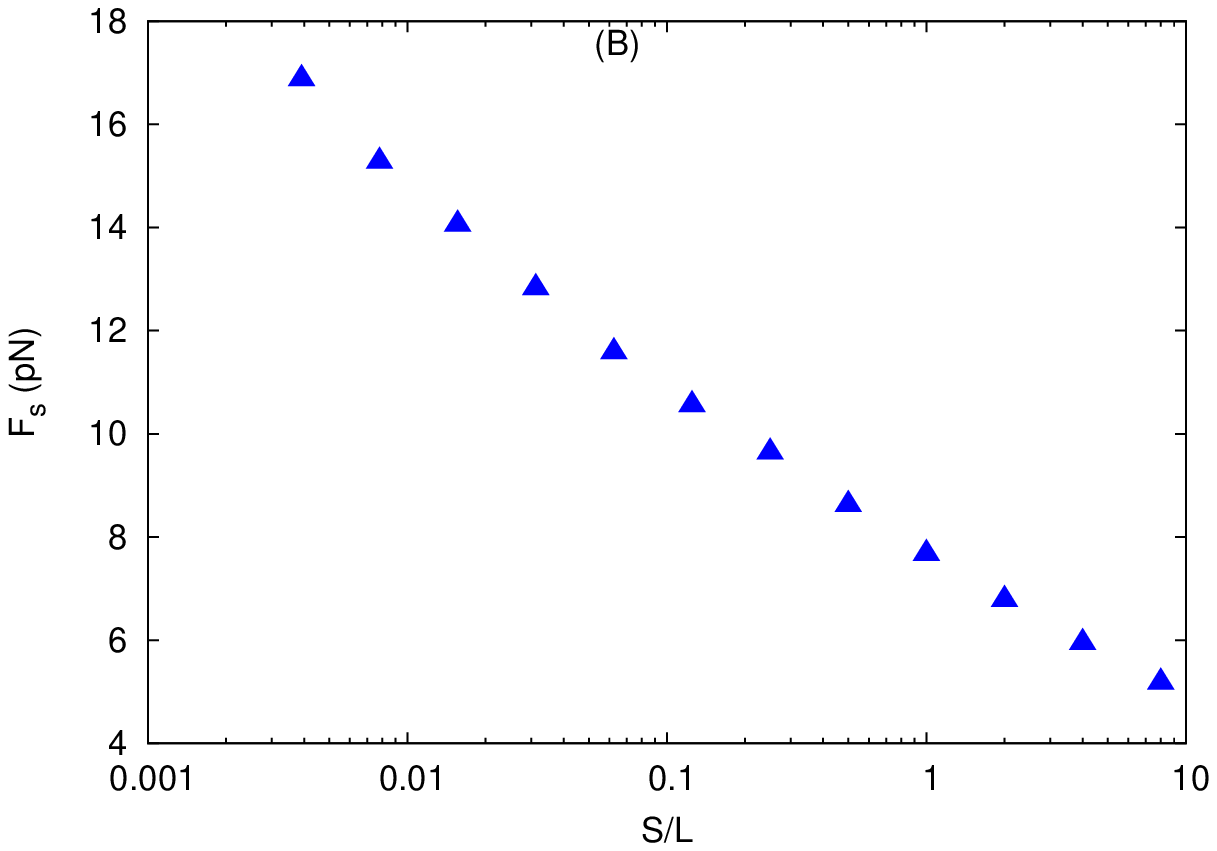}
\caption{Force-velocity characteristic for a single filament depends on the
relative time-scale between the filament and the barrier dynamics. (A): Velocity of
the barrier vs scaled force for different values of $S/L$. For large $S/L$,
the convex nature of force-velocity characteristic is lost.  
As $S/L$ increases, the  local fluctuations of the barrier become more important
and even for small $F$, the barrier velocity is 
not governed by the global movement alone, and hence $V$ does not decay
exponentially anymore. Here, we have used $L=64$.
(B): Stall force decreases as a function of $S/L$. Since local movements of the
barrier become more important for large $S/L$, the balance between global and
local movements is reached at a smaller force. Note however, that the $x$-axis
is plotted in a log-scale, indicating a weak dependence of stall force on the
time-scale. Here we have used $L=256$. The 
other parameters are same as in Fig. \ref{F_vs_V}.}
\label{F_vs_V_diff_m}
\end{figure}

We find the nature of the force-velocity curve depends on the relative time-scale 
of the barrier and filament dynamics. For faster barrier dynamics, the local
fluctuations of the barrier increases and as a result their contribution
to the net velocity is also higher. This means even for small force, the velocity is not 
dominated by the global movement (first term in Eq. \ref{eq:vf}) alone. In addition, our simple
expression for the contact probability $p_0=(1-W_0/U_0)$, which was derived neglecting the local fluctuations at the binding site, does not remain valid for fast barrier dynamics and $p_0$ increases with $F$ in this case (see our data in Fig. \ref{fig:p0m}). As a result, the velocity
does not decay exponentially for small force, but follows a slower decay. For a given value of F, in the small or intermediate range, as the barrier dynamics becomes faster, the velocity becomes higher and the convex nature of the curve is gradually lost. Moreover, since stalling phenomenon in 
our system can be described as a balance between global and local velocities of 
the barrier (see Eq. \ref{eq:vf}), larger contribution from local movement
implies this balance is reached at a smaller value of force. Therefore, for 
faster barrier dynamics we have a smaller stall force. We present our data in 
Figs. \ref{F_vs_V_diff_m}A and \ref{F_vs_V_diff_m}B. 

Our data in Fig. \ref{F_vs_V_diff_m}B imply that in the limit of infinitely 
slow barrier dynamics, when the barrier can be considered as an effectively 
rigid object, the stall force diverges. Note
that even in this limit, our model remains different from the rigid barrier case
studied in \cite{kirone}, where at least one filament is always bound to the 
barrier. For $N=1$ this would mean whenever there is a 
depolymerization, the barrier also moves down, along with the filament tip. 
On the contrary, we allow unsupported barrier in our system and when the barrier
is effectively rigid, it shows only global movement which is always in the 
upward direction. The force velocity curve is perfectly exponential in this case
and zero velocity is reached at $F \rightarrow \infty$ limit.


\subsection{Variation of the shape of the barrier with load}

\begin{figure}
\includegraphics[scale=0.8]{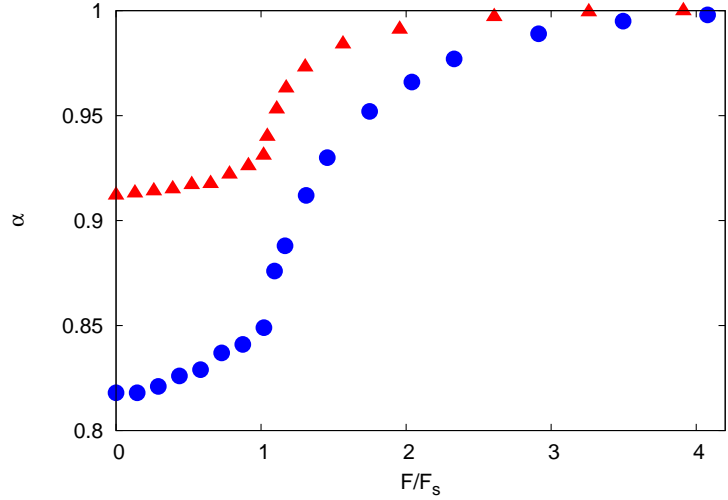}
\caption{Variation of $\alpha$ as a function of external load.
Close to the stalling force, $\alpha$ shows a sharp increase. Here, we have used
$S/L=1$ and $L=256$ (red triangle) and $128$ (blue circle). Other simulation
parameters are same as in Fig. \ref{F_vs_V}.}
\label{roughness}
\end{figure}

We have seen above how the barrier fluctuations affect the growth of the
filament. The barrier properties are also altered 
in this process. As the load increases, the height profile of the barrier 
shows larger variation across the system. We characterize it by measuring 
the scaling of average height with distance from the binding site:
$\langle h(r)-h(0) \rangle \sim r^{\alpha} $, where $h(r)$ is the height of a
site at a distance $r$ from the binding site. In Fig. \ref{roughness} we plot
$\alpha$ as a function of the external force, which 
shows that for small force $\alpha$ increases slowly, around the stalling force 
there is a sharp increase and finally for very large force, $\alpha$
saturates to unity. Note that large value of $\alpha$ indicates  
presence of large hills and valleys in the system. $\alpha=1$ corresponds to 
a phase separation of upslope and downslope bonds in the system which 
gives rise to one single large hill, the highest point being the binding site. 
This situation is similar to the case of an elastic membrane, when the membrane 
tension is large and the membrane is stretched.

\section{Results for multiple filaments}
\label{sec:multi}

In the case of $N$ filaments in the system, we mainly consider the case when 
the ratio $N/L$ is small. We assume the binding sites are uniformly 
placed on the lattice, at a distance $L/N$. Between the segment of two
successive binding sites, the same considerations as in a single filament case 
apply. We assume these segments are independent and apply our results
for the single filament case for each segment. 

To start with, we consider the velocity of the barrier due to its global movement 
$V_1= p_0 N d U_0 \exp(-\beta F d)$. As before, $p_0$ is the probability to find
a filament in contact with the barrier and $p_0N$ is the average number of 
bound filaments in the system. Here, we have neglected any correlation between
the binding sites. To calculate $p_0$, we write down master equations for 
average number $N_i$ of filaments at a distance $i$ from the corresponding 
binding sites. The steady state solutions of these equations can be obtained recursively for different values of $i$ (see appendix \ref{app:p0n} for details). 
For $N$ filaments we have  
\be
p_0 =\dfrac{(1-W_0/U_0)} {1+(N-1)\exp(-\beta F d)}.
\ee
For large $F$, the contact
probability becomes same as the single filament case. For small $F$, the contact 
probability is approximately $1/N$ times the single-filament value,
indicating that for small $F$, at most one filament is in contact 
with the barrier.

\begin{figure}
\includegraphics[scale=0.7]{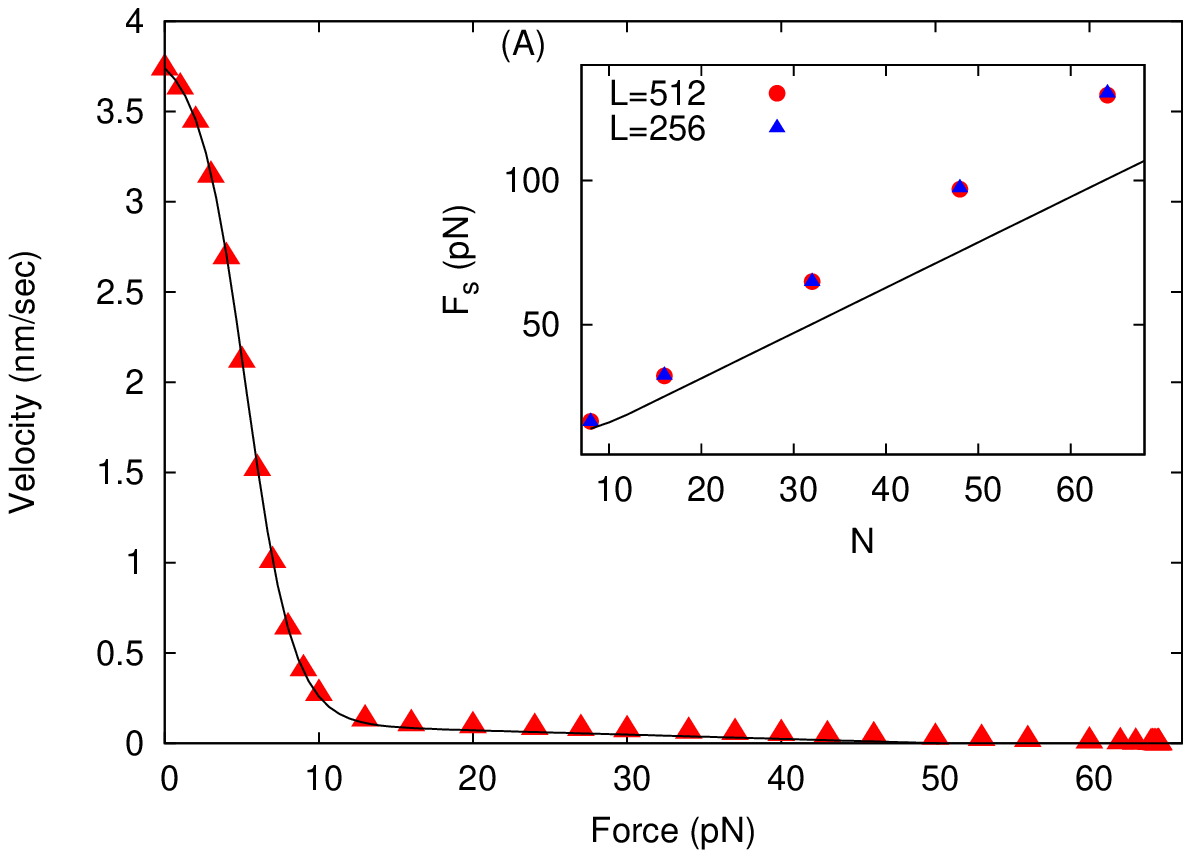}
\includegraphics[scale=0.7]{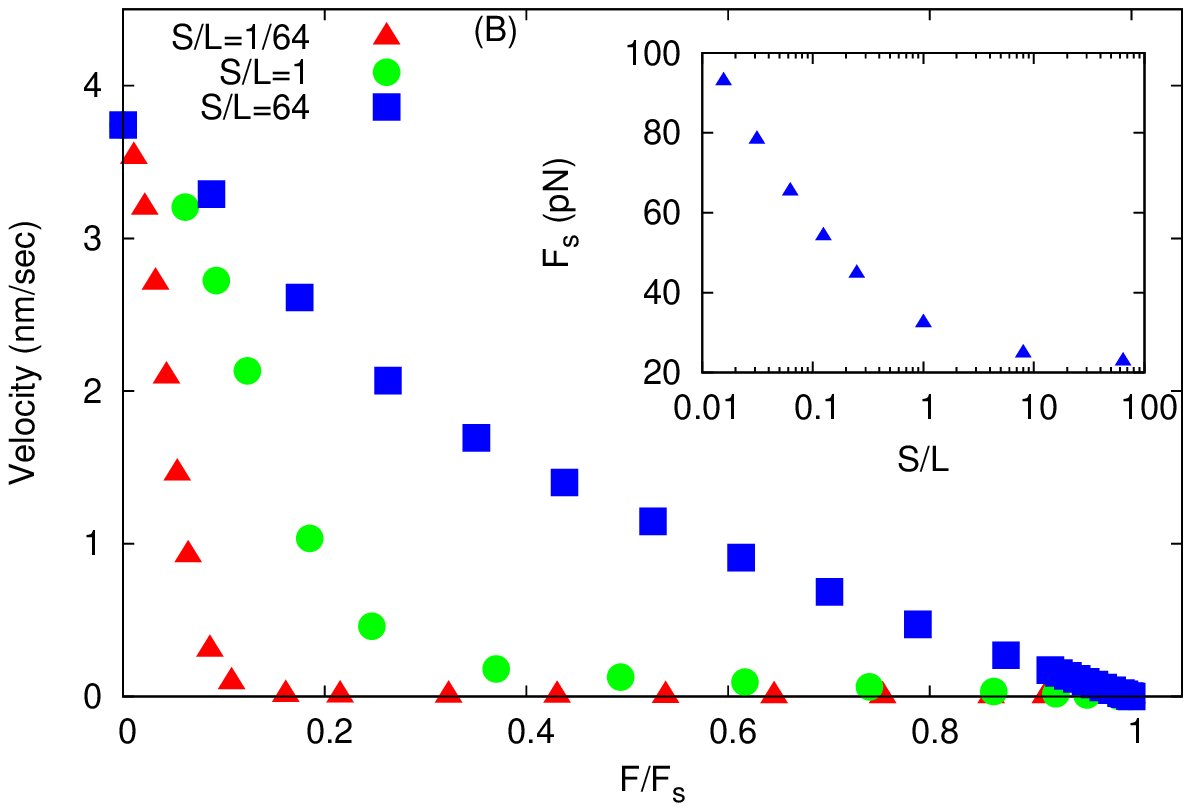}
\caption{Force-velocity characteristic for multiple filaments. (A): Velocity
shows very slow decay for large $F$, when global movement can be neglected and
$V$ can be assumed to be governed by local fluctuations alone. Here, we have
used $L=512$ and $N=32$. Inset shows stall force as a function of $N$ for two
different $L$ values. We find stall force scales linearly with $N$ and remains
independent of $L$. The continuous lines show analytical results. (B):
Dependence of  force-velocity characteristic on the time-scale of the barrier
dynamics. In this case we find same qualitative effect as in the single filament
case. Here, we have used $N=16$ and $L=128$.}
\label{large_N}
\end{figure}

For the local movement of the barrier, we need to calculate the probability to
find hills and valleys. As discussed above, for each segment between two
successive binding sites, we use our results for $p_v(i)$ and $p_h(i)$ for the
single filament case (with the modification that $i$ in this case varies from
$0$ to $(L/N-1)$). The velocity due to local fluctuations then becomes
\be
V_2=\frac{NdU_0}{L}\left [ \left (  p_v(0) (1+p_0) +  
\sum_{i=1}^{L/N-1} p_v(i) \right ) e^{-\beta F d /L} - (1-p_0)p_h(0) -
\sum_{i=1}^{L/N-1} p_h(i)  \right ]
\ee
The total velocity to leading order in $1/L$ and $N/L$ becomes
\be
V(F)=d \frac{(U_0-W_0)}{1+(N-1)e^{-\beta F d }}N~ e^{-\beta F d} + \frac{d
U_0N}{L}
\left [\left \{ p_v(0) (1+p_0)-p_h(0)(1-p_0) \right \} +
 \sum_{i=1}^{L/N-1} \left \{ p_v(i)\left ( 1-\frac{\beta F d}{L}
\right ) -p_h(i) \right \} \right ]
\label{eq:vn}
\ee
The stall force can be obtained by solving the above transcendental equation 
graphically for $V(F)=0$ and we compare the analytical stall force with our
simulation results in Fig. \ref{large_N}A inset. We find that the stall force
is independent of $L$ in this case and scales with $N$, which can be
easily seen from Eq. \ref{eq:vn}. Since the value of the stall force is rather large 
in this case, one can neglect global movement of the barrier close to the stalling regime. 
In addition, $p_0 \approx (1-W_0/U_0)$
for large force, and $\left ( p_v(i)-p_h(i) \right ) $ is of order $N/L$. 
Using these in Eq.
\ref{eq:vn} it directly follows  that the stall force for $N$ filaments is 
independent of $L$ and scales as $N$. 
We also investigate the effect of the time-scale of the barrier dynamics on the 
force-velocity dependence (Fig. \ref{large_N}B) and we find qualitatively the 
same effect as in $N=1$ case.

\section{Conclusions}
\label{sec:sum}

In this paper, we have studied force generation by a set of parallel filaments
polymerizing against a barrier. A similar question has been addressed in many 
recent works where the barrier was modeled as a rigid wall, which may have
a motion like a thermal ratchet 
\cite{peskin1993,mogilner1996,2mogilner1996,carlsson2000}, or may be a passive 
obstacle which can move only when pushed by the filaments 
\cite{kirone,ddas2014,catastroph1,hansda2014,mogilner1999,mogilner2003}. 
In this paper, we have considered a barrier with
thermal fluctuations but instead of modeling it as a rigid wall, we allow for
its shape fluctuations. In \cite{Baumgaertner2010} a similar aspect was studied where the barrier was modelled by a one dimensional Edwards-Wilkinson type membrane
under tension, which was being locally pushed by a set of growing filaments. The uncorrelated drive from the filaments gives rise to a KPZ type behavior in the correlated height fluctuations of the membrane, but this is associated with very slow crossover. Interestingly, the steady-state fluctuations of the driven membrane shows a non-monotonic behavior with the driving rate, where the strongly driven and weakly driven regimes are separated by a minimum in the width of the membrane profile. Although the filaments only impart local drive to the membrane, and 
no global movement of the membrane is considered in \cite{Baumgaertner2010}, the velocity  still shows an exponential dependence on the membrane tension, whereas in our model the exponential dependence is caused by the global movement and the local fluctuations generate a velocity that decreases roughly linearly with the external load.

One interesting result obtained in our system is the dependence of the
qualitative shape of the $F$-$V$ curve on the relative time-scale between the
filament polymerization and barrier fluctuation. For slow barrier dynamics, the
curve has a convex shape and $V$ shows an exponential decay for small and
moderate $F$. But for fast barrier dynamics when the local
fluctuations become more important, there is significant deviation from
exponential dependence. A similar effect was
reported in \cite{mogilner2012} for  a hybrid mesoscopic model that combines the 
microscopic dynamics of semi-flexible actin filaments 
and the viscous retrograde flow of actin network modeled as a macroscopic gel.     
It was shown that the force-velocity curve can be both convex and concave, 
depending on the characteristic time-scale of recoil of the gel-like network.
It is remarkable that our simple lattice gas model can reproduce this same
effect, which underlines the importance of the relative time-scale of obstacle
and filament dynamics on the force generation mechanism.

Throughout this paper,
we have considered the case  $\delta =d$, when the local movement of
the barrier occur in steps whose size is equal to that of a monomer. We have
verified (data not shown here) that many of our qualitative conclusions remain valid
for $\delta \ll d$. In other words, even when the shape fluctuations of the
barrier occur over much smaller length scales, their effect cannot be ignored.
We find that the stall force continues to show dependence on the barrier
properties. The relative time-scales between the filament and barrier dynamics
affects the $F-V$ curve in the same way. However, the quantitative
value of the stall force increases as smaller $\delta$ values are considered.

Finally, our simple model shows that a non-rigid obstacle can produce remarkable
effects on force generation of parallel actin filaments. Our results underline
the importance of the local shape distortions of an obstacle and indicate that
more research with detailed modeling of this aspect is required. 
Many of our conclusions are generic and can be expected to remain valid in 
systems where different descriptions of a non-rigid obstacle are used. This also 
opens up the possibility of observing some of these effects in experiment. For example, 
the change of shape of the barrier with external load can be monitored in an experiment
and our prediction that the height variation across the barrier increases with load, can be 
explicitly verified. The key feature of a fluctuating barrier is that one component of velocity
comes from the local fluctuations and a direct measurement of this component will surely give insights into the effects of barrier fluctuations. Our model shows that for multiple filaments close to stalling regime, velocity is dominated by these local movements and we also predict 
the scaling behavior of this velocity with filament density and barrier size. It would be interesting to verify these predictions in experiments, which would not only shed light on the qualitative nature of the local fluctuations but would also provide insights about their quantitative behavior. 

\section{Acknowledgements} 
The computational facility used in this work was 
provided through Thematic Unit of Excellence on Computational Materials Science, 
funded by Nanomission, Department of Science and Technology, India.


\appendix


\section{Calculation of $p_v(i)$ and $p_h(i)$ for single filament}
\label{app:toy}
\renewcommand{\theequation}{A-\arabic{equation}}
\setcounter{equation}{0}
\renewcommand{\thefigure}{A-\arabic{figure}}
\setcounter{figure}{0}

\begin{figure}
\includegraphics[scale=0.7]{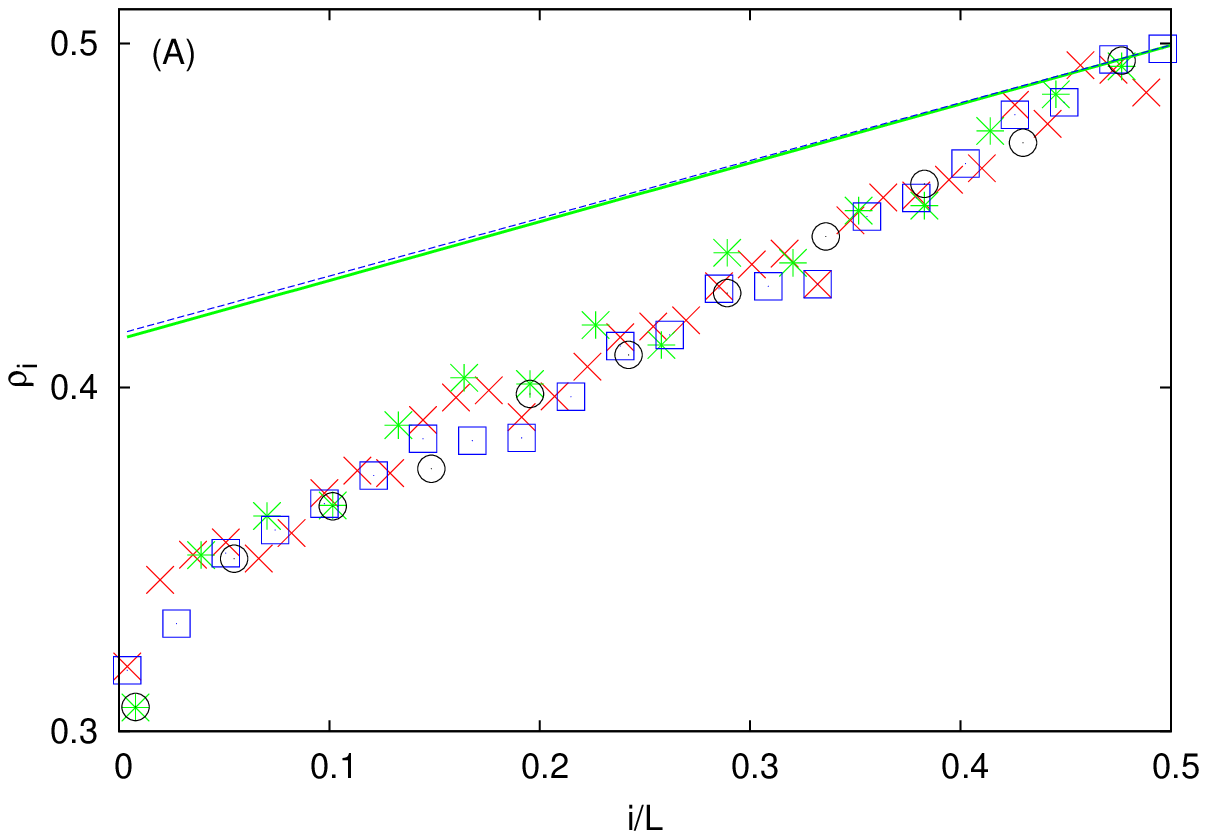}
\includegraphics[scale=0.7]{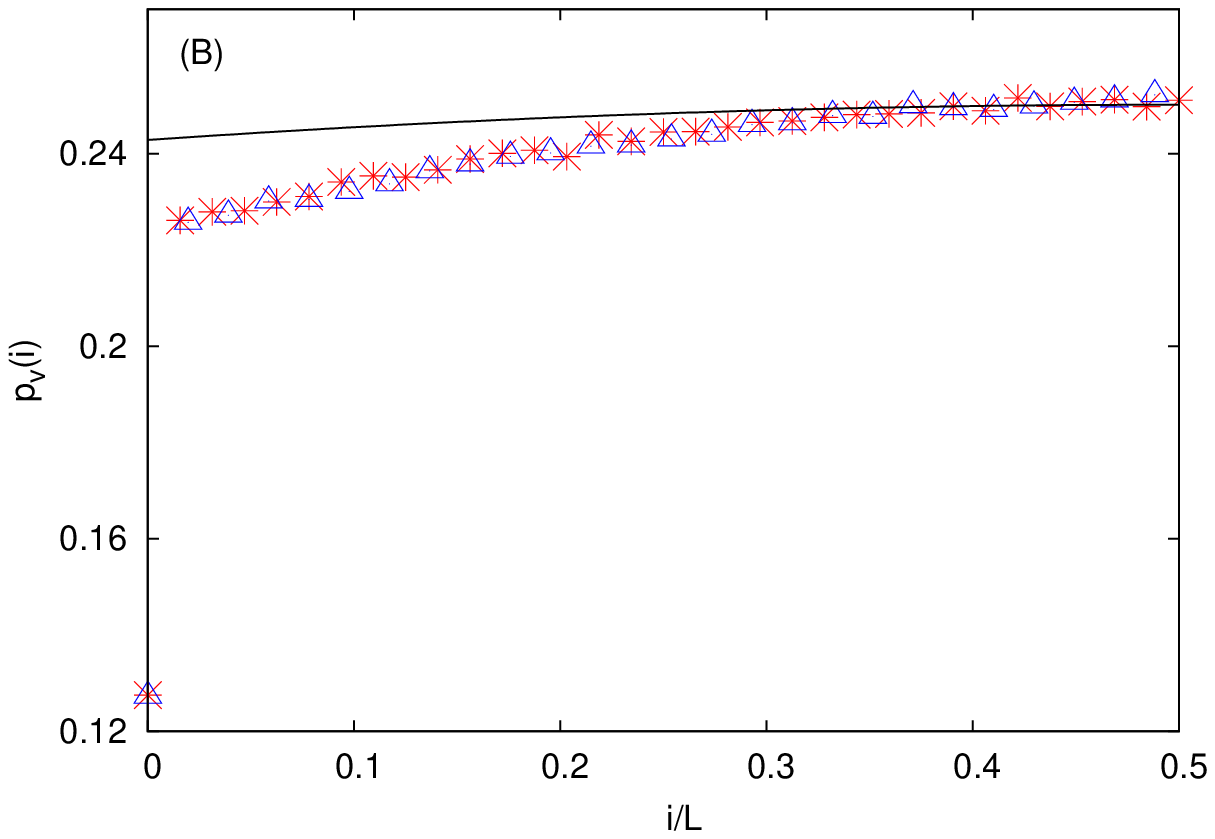}
\includegraphics[scale=0.7]{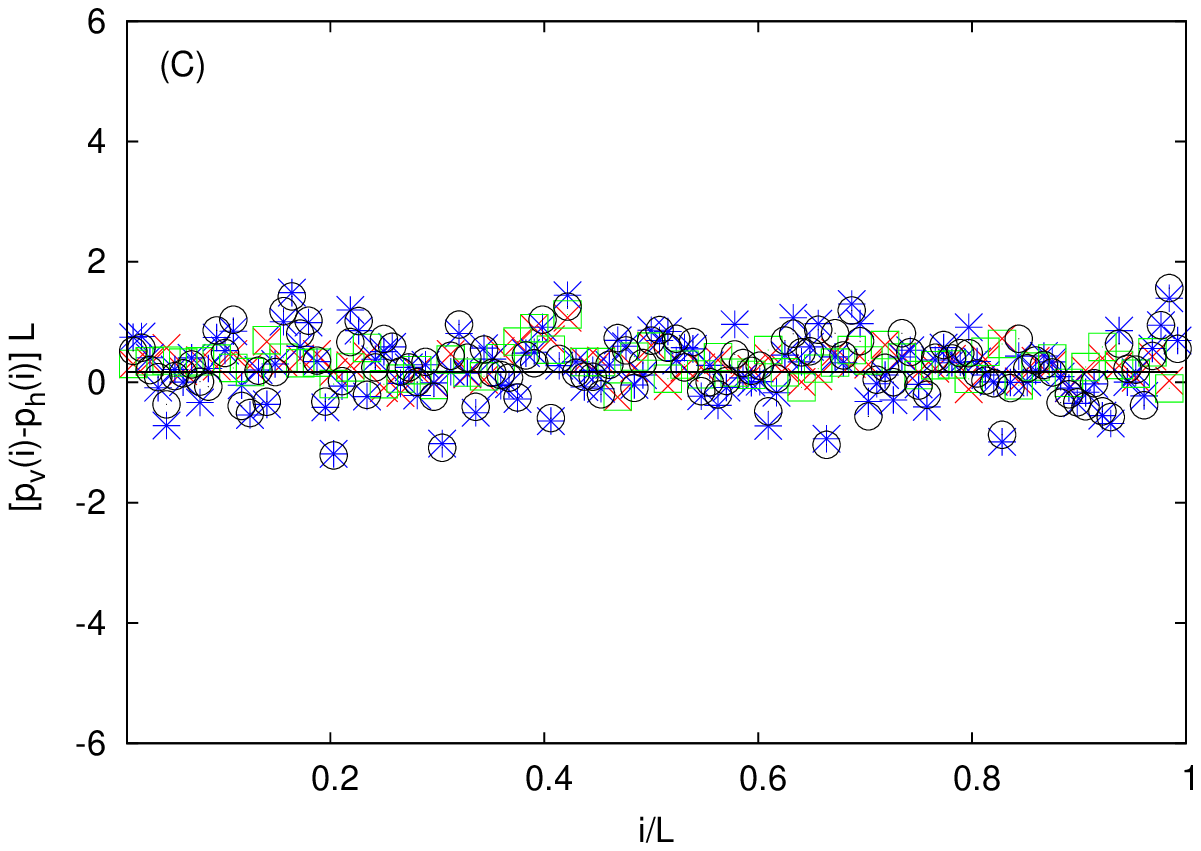}
\includegraphics[scale=0.7]{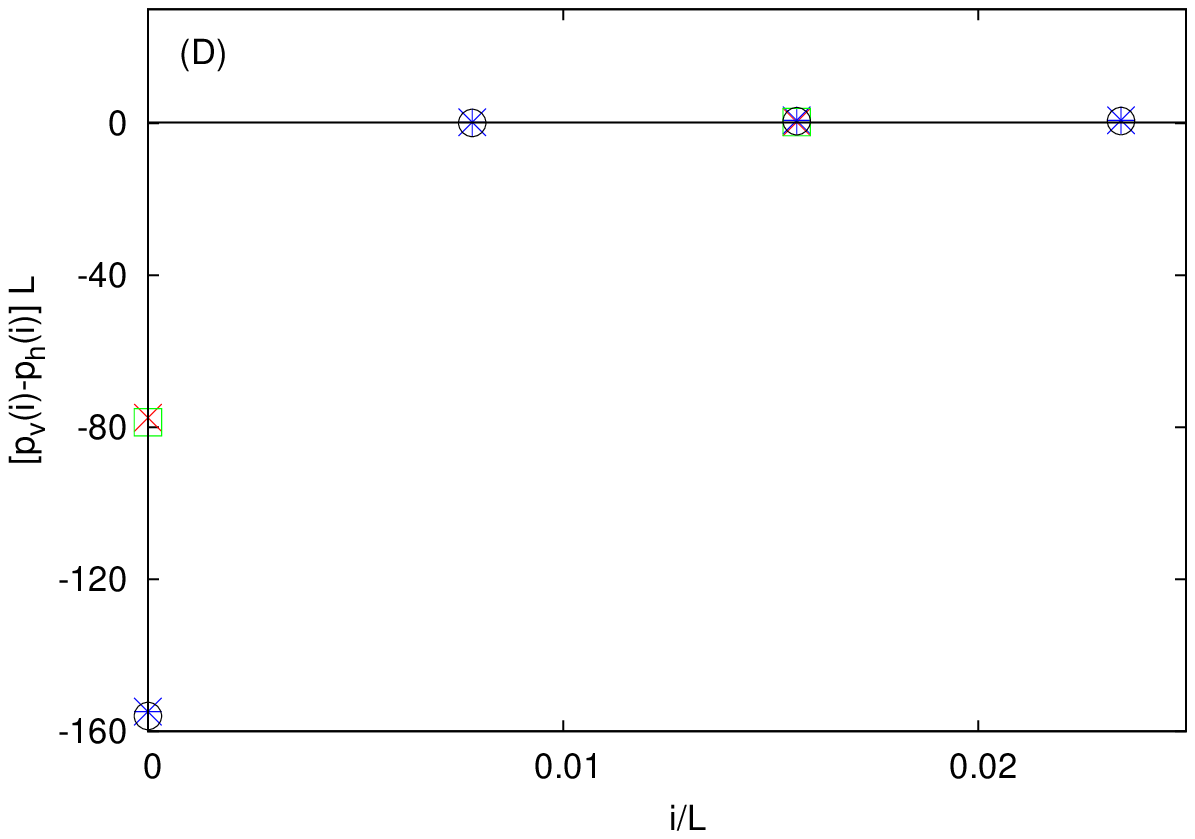} 
\caption{Average shape of the barrier for single filament. Discrete points show
simulation results and continuous lines show analytical predictions. (A): Probability
$\rho_i$ to find an upslope bond as a function of scaled distance $i/L$ from the
binding site. $\rho_i=1/2$ for $i=L/2$ and for larger $i$, we have
$\rho_i = 1-\rho_{i-L/2}$. The open symbols correspond to $F=0$ and the close symbols
correspond to $F=4 pN$. Symbols $\ast$ and $\circ$ are for $L=128$ and $\times$
 and $\Box$ are for $L=256$. These data show that, except close to the binding
site, $\rho_i$ increases linearly with $i$ with a gradient $\sim 1/L$. We also
find that $\rho_i$ remains almost same for these $F$ values. The continuous
lines are analytical predictions, where green solid line is for $F=0$ and blue
dashed line is for $F=4 pN$. (B): Probability $p_v(i)$
to find a valley at a distance $i$ from the binding site. For $i=0$ the
probability is substantially smaller compared to the rest of the system, which
means it is rather unlikely to find a valley at the binding site. The
symbols $\ast$ and $\Delta$ represent $F=0 pN$ and $4 pN$, respectively. We have
used $L=512$ here. (C) and (D): $[p_v(i) - p_h(i)]$ shows a
sharp jump at $i=0$ and then remains constant at a value that scales as $1/L$. 
The open symbols correspond to $F=0$ and the closed symbols
correspond to $F=4 pN$. Symbols $\ast$ and $\circ$ are for $L=256$ and $\times$
are $\Box$ are for $L=512$. }
\label{N_1_density_profile}
\end{figure}

The shape of the barrier changes due to transition between local hills and
valleys. The probability to find a hill at a site $s$ located at a distance $i$ from
the binding site is $p_h(i) $ and it can be written as $\rho_i (1-\rho_{i+1})$,
where $\rho_i$ is the probability that the bond preceding the site $s$ has $\pi/4$
orientation and  $ (1-\rho_{i+1})$ is the probability that the bond
immediately after the site  $s$ has $-\pi /4$ orientation. Here, we have used
mean-field theory and neglected correlation between the bonds. The probability
to find a valley at site $s$ can similarly be written as $(1-\rho_i)
\rho_{i+1}$. The transition rate from a hill to a valley is $R_-$ and the
reverse process occurs with rate $R_+$. For $i \neq 0$, $R_+/R_- = \exp (-\beta
F d /L)$. However, when $i=0$, or, in other words, the site $s$ is the binding
site itself, then although valley to hill transition is not affected, the
reverse transition can take place only when the filament is not in contact with
the binding site. We therefore make the simplifying assumption that the 
effect of the filament can be included by merely rescaling the  hill to valley
transition rate at the binding site by the probability that the filament
is in contact. In section \ref{sec:single} we calculate the contact probability  
$p_0=1-W_0/U_0 \simeq 1/2$. The master equations describing the time-evolution of 
$\rho_i$ can then be written as
\be
\frac{d\rho_i}{dt}=(1-\rho_i)( R_-\rho_{i-1} + R_+ \rho_{i+1} )- \rho_{i} [R_-
(1-\rho_{i+1})+ R_+(1-\rho_{i-1})], \;\;\;\;\;\;\mbox{for $2 \leq i \leq L-1$}
\label{eq:ri}
\ee
and at the binding site, 
\be 
\frac{d\rho_1}{dt}=(1-\rho_1) [ R_-(1-p_0) \rho_{L} +  R_+ \rho_{2} ]- 
\rho_{1} [R_- (1-\rho_{2})+ R_+\rho_{1} (1-\rho_{L})], 
\label{eq:r1}
\ee      
where we have applied periodic boundary condition, which also gives
\be
\frac{d\rho_L}{dt}=(1-\rho_L)( R_-\rho_{L-1} + R_+ \rho_{1} )- \rho_{L} [R_-
(1-\rho_{1})(1-p_0)+ R_+(1-\rho_{L-1})].
\label{eq:rl}
\ee    
We solve the above equations in the steady state when the left hand sides vanish. To
leading order in $1/L$, we find $\rho_i = a+bi/L$, where $a$ and $b$ are related 
via the condition $\sum_{i=1}^{L} \rho_i = L/2$ and $b$ satisfies the quadratic 
equation 
\be
\left [ \frac{\beta F d}{2L}-\frac{p_0}{4}\Big(1-\frac{2}{L}\Big)
\right ]b^2 + \left [1-\frac{\beta F d}{4L}-\frac{p_0}{2}\Big(1-\frac{1}{L}\Big) 
\right ]b +\frac{1}{4}\Big(\frac{\beta F d}{L}-p_0\Big) =0,
\ee
one of whose roots can be discarded from the condition that $0 \leq \rho_i \leq 1$ 
for all $i$. 
For a given $F$, therefore, $\rho_i$ varies linearly with the distance from the
binding site with a gradient $1/L$. For $F=0$, we have 
$a=(\sqrt{2}-1)$ and $b=(3-2\sqrt{2})$. 
For $0 \leq F \leq F_s$, the range of variation of
$a$ and $b$ are rather small and occur at third or higher decimal places. 
Therefore, $\rho_i$ does not change significantly with $F$. Our simulation data
in Fig. \ref{N_1_density_profile}A show similar qualitative behavior, although
close to the binding site there is deviation of $\rho_i$ from linearity. The
quantitative values of $a$ and $b$ however, do not match with simulations. We
attribute this mismatch to the mean field theoretic assumptions used in our
calculation.

We calculate $p_v(i)$ and $p_h(i)$ from $\rho_i$ and compare with
simulation in Fig. \ref{N_1_density_profile}B. Notice that from our analytical
expression for $\rho_i$, it follows immediately  that $\left ( p_v(i) -
p_h(i)\right ) $ is 
independent of $i$ and $\sim b/L$. This has important consequence for our
calculation of $V_2$ in section \ref{sec:single}. Moreover, 
the probability that the filament is in contact with a valley is
given by $p_v(0) p_0$ and our numerical results in Fig. \ref{N_1_density_profile}B
show that this probability is rather small.   

\section{Variation of contact probability for a single filament with load for fast and slow barrier dynamics}
\label{app:p0}
\renewcommand{\theequation}{B-\arabic{equation}}
\setcounter{equation}{0}
\renewcommand{\thefigure}{B-\arabic{figure}}
\setcounter{figure}{0}
\begin{figure}
\includegraphics[scale=0.8]{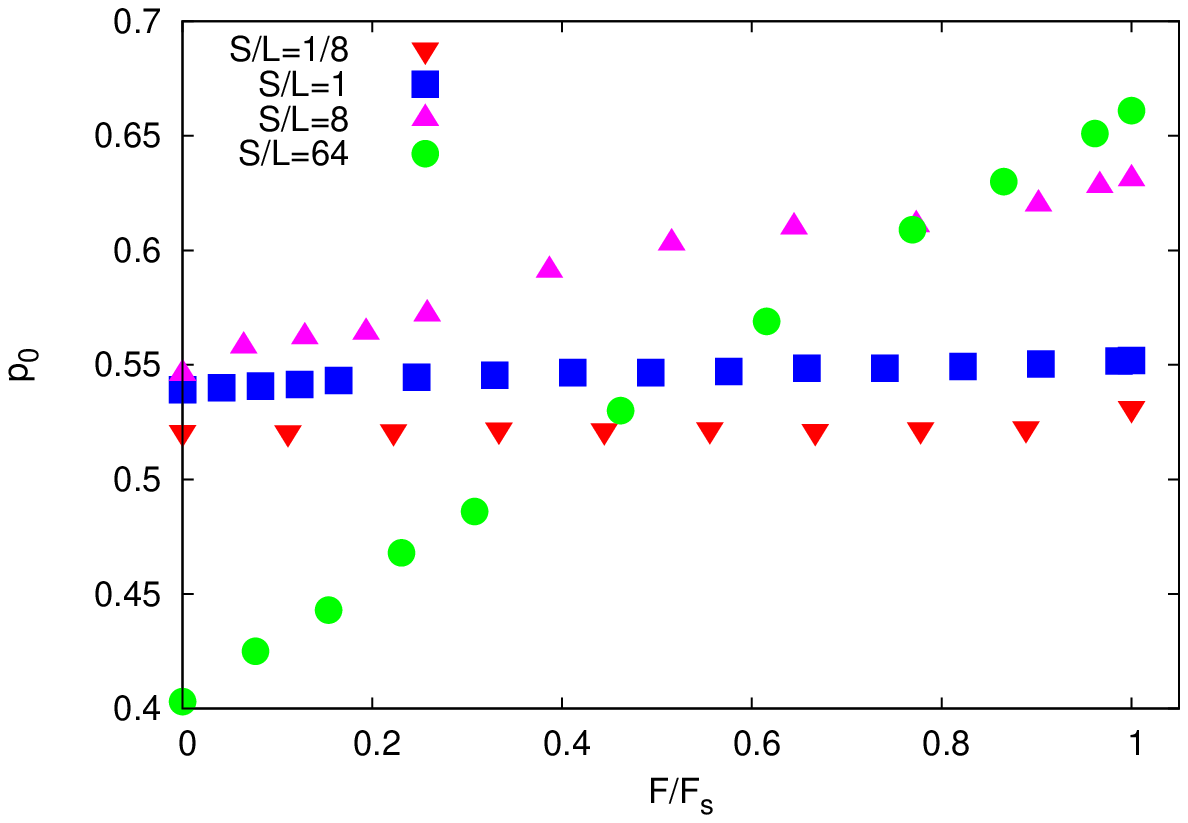}[ht] 
\caption{Contact probability $p_0$ as a function of $F$ for single filament. Our analytical 
calculation yields $p_0 = (1-W_0/U_0) \simeq 0.5$. For slow barrier dynamics, we find reasonable
agreement. But for fast barrier dynamics, our analytical prediction does not remain valid anymore and $p_0$ increases with $F$. The simulation parameters are as in Fig. \ref{F_vs_V}.}
\label{fig:p0m}
\end{figure}


\section{Calculation of contact probability for multiple filaments}
\label{app:p0n}
\renewcommand{\theequation}{C-\arabic{equation}}
\setcounter{equation}{0}
\renewcommand{\thefigure}{C-\arabic{figure}}
\setcounter{figure}{0}

Let $N_i$ be the average number of filaments at a distance $i$ from the
respective binding sites. By definition, $N_0$ is the average number of bound
filaments and the contact probability is $p_0 = N_0/N$. The time-evolution
equations for $N_i$ can be written as 
\bea
\frac{dN_0}{dt}&=&U_0N_{1}-\{(N_0-1)U_0 e^{-\beta F d} +W_0\}N_0, \label{eq:n0} \\
\frac{dN_1}{dt}&=&\{(N_0-1)U_0 e^{-\beta F d}+W_0\}N_{0}+U_0N_{2}-(N_0U_0 
e^{-\beta F d}+W_0+U_0)N_1 \label{eq:n1}, \\
\frac{dN_i}{dt}&=&(N_0U_0 e^{-\beta F d}+W_0)N_{i-1}+U_0N_{i+1}-(N_0U_0 
e^{-\beta F d}+W_0+U_0)N_i \label{eq:ni} 
\;\;\;\; \mbox{for $i \geq 2$}.
\eea   
Here, we have assumed that the distance $i$ between the filament tip and the
binding site can change only due to polymerization and depolymerization dynamics
and the global movement of the whole barrier due to polymerization of bound
filaments. We have neglected local height fluctuations occurring at the binding 
sites. As we show
below, this approximation works reasonably well as long as the filament density
$N/L$ is small and the time-scale of barrier fluctuation is comparable to, or
slower than the filament dynamics. For very fast
motion of the barrier, the height fluctuations at the binding sites become more
frequent and this assumption breaks down.

\begin{figure}
\includegraphics[scale=0.7]{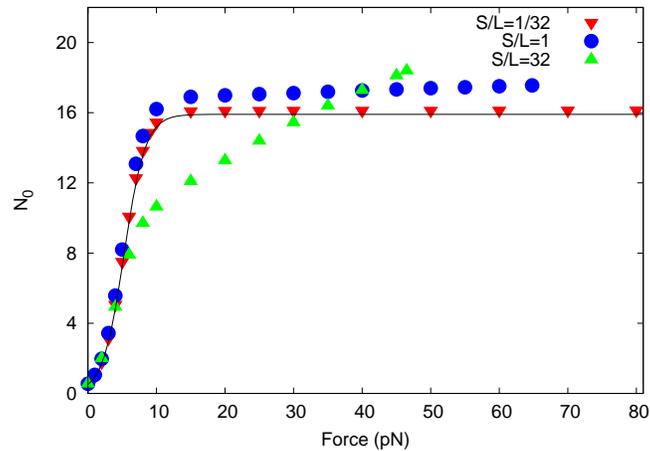}
\caption{Average number of bound filaments $N_0$ as a function of force $F$. For
slow barrier dynamics, our analytical prediction in Eq. \ref{eq:bound} agree well
with numerics. But as the barrier dynamics becomes faster, deviations are
observed. Here we have used $L=256$, $N=32$. Other simulation parameters are
same as in Fig. \ref{F_vs_V}. }
\label{fig:nb}
\end{figure}

Solving the Eqs. \ref{eq:n0}, \ref{eq:n1}, \ref{eq:ni} in the steady state, 
we obtain the recursion relation
\begin{equation}
N_{i+1}=\bigg(\frac{N_0U_0 e^{-\beta F d}+W_0}{U_0}\bigg)^{i}N_1; ~ i=1,2,...
\end{equation}
and
\begin{equation}
N_1=\frac{(N_0U_0 e^{-\beta F d}+W_0-U_0 e^{-\beta F d})}{U_0}N_0
\end{equation}
Using the normalization relation, $\sum N_i=N$ we get 
\begin{equation}
N_0=\frac{N(U_0-W_0)}{U_0-U_0 e^{-\beta F d}+NU_0 e^{-\beta F d}}
\label{eq:bound}
\end{equation}
and the contact probability has the form 
$p_0=\frac{(U_0-W_0)}{U_0+(N-1)U_0 e^{-\beta F d}}$. In Fig. \ref{fig:nb} 
we compare this result with simulation and find reasonable agreement.



\begin{thebibliography}{99}
\bibitem{review1} J. Howard, Mechanics of motor proteins and the cytoskeleton, 
Sunderland, MA: Sinauer Associates (2001).

\bibitem{review2} T. D. Pollard and J. A. Cooper, Actin, a central player in cell shape and movement, Science {\bf 326}, 1208 (2009).

\bibitem{review3} L. Blanchoin, R. B. Paterski, C. Sykes and J. Plastino, Actin dynamics, architecture and mechanics in cell motility, Physiol Rev {\bf 94}, 235 (2014). 

\bibitem{review4} P. Friedl and D. Gilmour, Collective cell migration in
morphogenesis, regeneration and cancer, Nat. Rev. Mol. Cell Biol. {\bf 10}, 445 (2009).

\bibitem{marcy2004}  Y. Marcy, J. Prost, M. F. Carlier, and C. Sykes, Forces
generated during actin-based propulsion: A direct measurement by
micromanipulation, Proc. Natl. Acad. Sci. U.S.A. {\bf 101}, 5992 (2004).

\bibitem{baudry2011} C. Brangbour, O. du Roure , E. Helfer, D. Démoulin, A.
Mazurier, M. Fermigier, M. F. Carlier, J. Bibette and J. Baudry, Force-velocity
measurements of a few growing actin filaments, PLoS Biol. {\bf 9}, e1000613 (2011). 

\bibitem{baudry2014}  D. Démoulin, M. F. Carlier, J. Bibette, and J. Baudry,
Power transduction of actin filaments ratcheting in vitro against a load,  Proc.
Natl. Acad. Sci. U.S.A. {\bf 111}, 17845 (2014).

\bibitem{theriot2005} S. H. Parekh, O. Chaudhuri, J. A. Theriot and D. A.
Fletcher, Loading history determines the velocity of actin-network growth, Nat.
Cell Biol. {\bf 7}, 1219 (2005).

\bibitem{mogilner2006}  M. Prass, K. Jacobson, A. Mogilner and M. Radmacher,
Direct measurement of the lamellipodial protrusive force in a migrating cell, J.
Cell. Biol. {\bf 174}, 767 (2006). 
\bibitem{zimm} J. Zimmermann, C. Brunner, M. Enculescu, M. Goegler, A.
Ehrlicher, J. K{\"a}s  and M. Falcke, Actin filament elasticity and retrograde
flow shape the force-velocity relation of motile cells, Biophys. J. {\bf 102}, 287 (2012).

\bibitem{theriot2003} P. A. Giardini, D. A. Fletcher and J. A. Theriot, Compression forces generated by actin comet tails on lipid vesicles,  Proc. Natl. Acad. Sci. U.S.A. {\bf 100}, 6493 (2003).

\bibitem{theriot2007} M. J. Footer, J. W. J. Kerssemakers, J. A. Theriot and M. Dogterom, Direct measurement of force generation by actin filament polymerization using an optical trap,  Proc. Natl. Acad. Sci. U.S.A. {\bf 104}, 2181 (2007). 

\bibitem{pollard2004} D. R. Kovar and T. D. Pollard, Insertional assembly of actin filament barbed ends in association with formins produces piconewton forces,  Proc. Natl. Acad. Sci. U.S.A. {\bf 101}, 14725 (2004). 

\bibitem{peskin1993} C. S. Peskin, G. M. Odell and G. F. Oster, Cellular motions
and thermal fluctuations: the Brownian ratchet, Biophys. J. {\bf 65}, 316 (1993).

\bibitem{carlsson2000} A. E. Carlsson, Force–velocity relation for growing
biopolymers, Phys. Rev. E {\bf 62}, 7082 (2000).


\bibitem{jphys2006} N. J.  Burroughs and D. Marenduzzo, Growth of a semi-flexible 
polymer close to a fluctuating obstacle:
application to cytoskeletal actin fibres and testing of ratchet models, J. Phys.: Condens. Matter {\bf 18}, S357 (2006).

\bibitem{schaus} T. E. Schaus, G. G. Borisy, Performance of a population of
independent filaments in lamellipodial protrusion, Biophys. J. {\bf 95}, 1393 (2008).

\bibitem{kirone} K. Tsekouras, D. Lacoste, K. Mallick and J. F. Joanny, Condensation of actin filaments pushing against a barrier, New J. Phys. {\bf 13}, 103032 (2011).

\bibitem{krawczyk2011} J. Krawczyk and J. Kierfeld, Stall force of polymerizing microtubules and filament bundles, Europhys. Lett. {\bf 93}, 28006 (2011).

\bibitem{ddas2014} D. Das, D. Das and R. Padinhateeri, Collective force generated by multiple biofilaments can exceed the sum of forces due to individual ones, New J. Phys. {\bf 16}, 063032 (2014).


\bibitem{mogilner2012} J. Zhu and A. Mogilner, Mesoscopic model of actin-based
propulsion, PLoS Comp. Biol. {\bf 8}, e1002764 (2012). 

\bibitem{carlsson2014} R. Wang and A. E. Carlsson, Load sharing in the growth of bundled biopolymers, New J. Phys {\bf 16}, 113047 (2014).

\bibitem{hansda2014} D. K. Hansda, S. sen and R. Padinhateeri, Branching influences force-velocity curve and length fluctuations in actin networks, Phys. Rev. E {\bf 90}, 062718 (2014).

\bibitem{nirgov} N. S. Gov and A. Gopinathan, Dynamics of membranes driven by actin 
polymerization,  Biophys J {\bf 90}, 454 (2006).

\bibitem{atilgan} E. Atilgan, D. Wirtz and S. X. Sun, Mechanics and Dynamics of Actin-Driven thin membrane protrusions,  Biophys J {\bf 90}, 65 (2006).

\bibitem{liu} A. P. Liu, D. L. Richmond, L. Maibaum, S. Pronk, P. L. Geissler
and D. A. Fletcher, Membrane-induced bundling of actin filaments,  Nat. Phys. {\bf 4}, 789 (2008).

\bibitem{kpz} M. Kardar, G. Parisi and Y-C. Zhang, Dynamic scaling of growing 
interfaces, Phys. Rev. Lett. {\bf 56}, 889 (1986).


\bibitem{ew} S. F. Edwards and D. R. Wilkinson, The surface statistics of a 
granular aggregate, Proc. R. Soc. London {\bf 381}, 17 (1982).

\bibitem{pollard} T. D. Pollard, Rate constants for the reactions of ATP-and ADP-actin with the ends of actin filaments, J. Cell. Biol. {\bf 103}, 2747 (1986).


\bibitem{2mogilner1996} A. Mogilner and G. Oster, The physics of lamellipodial
protrusion, Euro. Biophys. J. {\bf 25}, 47 (1996).

\bibitem{mogilner1996} A. Mogilner and G. Oster, Cell motility driven by actin
polymerization, Biophys. J. {\bf 71},  3030 (1996).

\bibitem{catastroph1} D. Das, D. Das and R. Padinhateeri, Force induced
dynamical properties of multiple cytoskeletal filaments are distinct from that
of single filament, PLoS One {\bf 9} (12), e114014 (2014).



\bibitem{mogilner1999} A. Mogilner and G. Oster, The polymerizing ratchet model
explains the force velocity relation for growing microtubule, Eur. Biophys. J. {\bf 28}, 235 (1999).


\bibitem{mogilner2003} A. Mogilner and G. F. Oster, Force generation by actin
polymerization II: the elastic ratchet and tethered filaments, Biophys. J. {\bf 84}, 1591 (2003).

\bibitem{Baumgaertner2010} S. L. Narasimhan and A. Baumgaertner, Dynamics of a driven surface, J. Chem. Phys. {\bf 133}, 034702 (2010).

\end{thebibliography}
\end{document}